\documentstyle[aps]{revtex}

\begin{document}
\author{M. Ruzzi}
\address{mruzzi@fsc.ufsc.br - Depto.de F\'{\i}sica\\
Universidade Federal de Santa Catarina\\
88040 - 900 Florian\'{o}polis, S.C., Brazil}
\author{D. Galetti}
\address{Instituto de F\'{\i}sica Te\'{o}rica\\
Universidade Estadual Paulista\\
Rua Pamplona 145\\
01405 - 900 S\~{a}o Paulo, S.P., Brazil}
\title{Schwinger, Pegg and Barnett approaches and a relationship between angular
and Cartesian quantum descriptions II: Phase Spaces.}
\maketitle

\begin{abstract}
Following the discussion -- in state space language -- presented in a
preceding paper, we work on the passage from the phase space description of
a degree of freedom described by a finite number of states (without
classical counterpart) to one described by an infinite (and continuously
labeled) number of states. With that it is possible to relate an original
Schwinger idea to the Pegg and Barnett approach to the phase problem. In
phase space language, this discussion shows that one can obtain the
Weyl-Wigner formalism, for both Cartesian {\em and} angular coordinates, as
limiting elements of the discrete phase space formalism.

\medskip\ 

\end{abstract}

\section{Introduction}

In a previous work it has been shown that the usual quantum descriptions of
Cartesian and angular coordinates in state space can both be seen as
different limiting cases of the Schwinger program of treating quantum
discrete variables \cite{eu}. The limiting process involved reproduces the
Pegg and Barnett approach to phase variables in the case of angle/angular
momentum variables\cite{pegg}. The purpose of this work is to translate that
discussion into a phase space point of view, which might be a way to unify
under a same structure three apparently different formalisms, each one
adapted to one specific kind of quantum variable, namely Cartesian, angular
and discrete. In doing so, we again relate the Pegg and Barnett and
Schwinger approaches, now through the phase space representatives of number
and phase operators.

The phase-space picture is a well established picture of quantum mechanics 
\cite{weyl,wigner,moyal,hillery,balazs,leaf,degroot,kim}, specially if one
deals with degrees of freedom with classical counterpart, a situation in
which the Weyl-Wigner formalism is the undisputed approach. Nevertheless, to
cope with variables such as rotation angle and angular momentum, the
formalism had to be adapted in order to account for the inherent periodicity
involved. This was accomplished in the late seventies \cite{berry,muk} and
developed to full extent in \cite{biza}. However, when one deals with
degrees of freedom {\it without }classical counterpart, the formulations
discussed above are not applicable. In these cases, there is a formalism
completely capable to deal with the peculiarities of the finite/discrete
character of the variables \cite{gapi1,gapi2,ruga}, much in the same spirit
of the Weyl-Wigner formalism itself. Following the procedure shown in \cite
{eu} which led from discrete variables to the cases with classical
counterpart, we show that from the discrete phase space formalism naturally
emerges the usual Weyl-Wigner formalism and {\it also} its rotation
angle-angular momentum version. In addition, some properties of the discrete
Wigner function are also discussed. From a more rigorous mathematical point
of view, these limiting processes presented in this context have been also
discussed in Refs. \cite{bar1,bar2,bar3}.

This paper is organized as follows. In section II we briefly present the
main ideas of the discrete phase space representation drawing attention to
some properties of the discrete Wigner function, while in section III we
discuss the limiting processes which lead the original operator bases to the
well-known Weyl-Wigner continuous case as well as the particular case of
rotations. Finaly, section IV is devoted to the conclusions.

\section{Discrete Phase Space}

As has been already shown in the past\cite{gapi1,gapi2,ruga}, a discrete
phase space representation of a quantum mechanical degree of freedom which
is characterized by a finite number of states, therefore with no classical
counterpart, can be established if we are given a basis in the corresponding
operator space. One such basis has been introduced by Schwinger in his
seminal paper on this subject\cite{schw}, constructing it out of some
particular unitary cyclically shifting pairs of operators, and another one
has been proposed that basically considers the double Fourier transform of
that of Schwinger\cite{gapi1}. As was previously shown, once we are provided
with such an operator basis, it is a direct task to obtain the discrete
phase space representatives of the operators acting on the state space from
which we started. To briefly summarize those results let us consider the
operator basis and recall its main properties.

The discrete phase space formalism is set over the basis elements 
\begin{equation}
G(j,l)=\frac{1}{N}\sum_{m,n=-h}^{h}U^{m}V^{n}\exp \left( \frac{i\pi mn}{N}%
\right) \exp \left[ -\frac{2\pi i}{N}\left( mj+nl\right) \right] \exp \left[
i\pi \phi \left( m+h,n+h;N\right) \right] ,  \label{14}
\end{equation}
where $(j,l)\in \lbrack -h,h],$ $h=\frac{N-1}{2}$ (for simplicity, odd $%
N^{\prime }$s will be considered, as even values only require only a little
more care and a heavier notation). The modular phase $\phi \left(
m,n;N\right) $, included to warrant an explicit mod $N$ symmetry in the
summing indices of the basis, is given by 
\begin{equation}
\phi \left( m,n;N\right) =NI_{m}^{N}I_{n}^{N}-mI_{n}^{N}-nI_{m}^{N}
\label{15}
\end{equation}
with 
\begin{equation}
I_{k}^{N}=\left[ \frac{k}{N}\right]  \label{16}
\end{equation}
standing for the integral part of $k$ with respect to $N$. The $U$'s and $V$%
''s are the Schwinger unitary operators\cite{schw}, shortly reviewed in \cite
{eu}.

As a basis, the set (\ref{14}) can be used to represent all linear operators
acting on the given $N$-dimensional state space; this can be accomplished by
a direct decomposition 
\begin{equation}
\hat{O}=\sum_{m,n=0}^{N-1}O\left( m,n\right) G\left( m,n\right) ,  \label{12}
\end{equation}
where the coefficient, $O\left( m,n\right) $, that gives rise to the
representative of the operator $\hat{O}$ in the discrete phase space\cite
{gapi1}, is given by 
\begin{equation}
O\left( m,n\right) =\frac 1NTr\left[ G\left( m,n\right) \hat{O}\right] ,
\label{13}
\end{equation}
where we used the fact that $G\left( m,n\right) $ is self adjoint.

The basic properties of the basis, Eq. (\ref{14}), are 
\begin{equation}
1)\;\;Tr\left[ G\left( m,n\right) \right] =1;  \label{17}
\end{equation}
\begin{equation}
2)\;\;Tr\left[ G^{\dagger }\left( m,n\right) G\left( r,s\right) \right]
=N\;\delta _{m,r}^{\left[ N\right] }\;\delta _{n,s}^{\left[ N\right] };
\label{18}
\end{equation}
\[
3)\;\;Tr\left[ G^{\dagger }\left( m,n\right) G\left( u,v\right) G\left(
r,s\right) \right] =\sum_{a,b,c,d=-h}^{h}\frac{1}{N^{2}}\exp \left[ \frac{%
i\pi }{N}\left( bc-ad\right) \right] 
\]
\begin{equation}
e^{\left[ -i\pi \phi \left( a+c+h,b+d+h;N\right) \right] }\exp \left\{ \frac{%
2\pi i}{N}\left[ a\left( m-u\right) +b\left( n-v\right) +c\left( m-r\right)
+d\left( n-s\right) \right] \right\} ,  \label{20}
\end{equation}
where the last expression is important for the mapping of products of
operators\cite{gama}. Particular interest resides in the mapping of the
commutator of two operators, for then it is possible to study, for example,
the time evolution of the density operator in the von Neumann-Liouville
equation\cite{ruga,garu}.

\subsection{The Discrete Wigner function}

The phase space representative of the density operator in the discrete
approach is also referred to as (discrete) Wigner function\cite
{gapi1,wooters,cohen}. If the (pure) state of a given system is described by 
\begin{equation}
|\psi \rangle =\sum_{n}\psi _{n}|u_{n}\rangle ,  \label{p1}
\end{equation}
where $\left\{ |u_{n}\rangle \right\} $ is the (complete and orthonormal)
set of eigenvectors of the Schwinger operator $U$, then the use of eq.(\ref
{13}) leads to a Wigner function of the form 
\begin{equation}
\rho _{w}(m,n)=\frac{1}{N^{2}}\sum_{j,l,k}\psi _{k}^{\ast }\psi _{k-l}\exp %
\left[ \frac{2\pi i}{N}\left( jk-\frac{jl}{2}-mj-nl\right) \right] ,
\label{p2}
\end{equation}
or 
\[
\rho _{w}(m,n)=\frac{1}{N^{2}}\sum_{l,k}\psi _{k}^{\ast }\psi _{k-l}\frac{%
\sin \left[ \pi (k-m-l/2)\right] }{\sin \left[ \frac{\pi }{N}(k-m-l/2)\right]
}\exp \left[ -\frac{2\pi i}{N}nl\right] . 
\]
Its main properties are, in direct analogy with the usual continuous Wigner
function:

\begin{itemize}
\item[1)]  It is a real function, as it follows from the hermicity of the
basis elements.

\item[2)]  Summing it over each one of its indices gives the probability
distribution in the other. For example: 
\begin{equation}
\sum_{n}\rho _{w}(m,n)=\sum_{n}\frac{1}{N^{2}}\sum_{j,l,k}\psi _{k}^{\ast
}\psi _{k-l}\exp \left[ \frac{2\pi i}{N}\left( jk-\frac{jl}{2}-mj-nl\right) %
\right] ,  \label{i1}
\end{equation}
such that 
\begin{equation}
\sum_{n}\rho _{w}(m,n)=\frac{1}{N}\sum_{j,l,k}\psi _{k}^{\ast }\psi
_{k-l}\exp \left[ \frac{2\pi i}{N}\left( jk-\frac{jl}{2}-mj\right) \right]
\delta _{l,0}^{[N]},  \label{i2}
\end{equation}
and so 
\begin{equation}
\sum_{n}\rho _{w}(m,n)=|\psi _{m}|^{2}.  \label{i5}
\end{equation}
And in the same way, the summation over $\left\{ m\right\} $ would lead to
the probability distribution associated to the eigenstates of the Schwinger
operator $V$.

\item[3)]  It must be different from zero in at least $N$ sites in the
discrete phase space. Writing it as 
\begin{equation}
\rho _{w}(m,n)=\frac{1}{N^{2}}\sum_{j,l}\exp \left[ -\frac{2\pi i}{N}\left(
mj+nl\right) \right] \sum_{k}\psi _{k}^{\ast }\psi _{k-l}\exp \left[ \frac{%
2\pi i}{N}\left( jk-\frac{jl}{2}\right) \right] ,  \label{i7}
\end{equation}
it is clear that it is the double Fourier transform of the quantity $\rho
_{s}(j,l)$%
\begin{equation}
\rho _{s}(j,l)=\sum_{k}\psi _{k}^{\ast }\psi _{k-l}\exp \left[ \frac{2\pi i}{%
N}\left( jk-\frac{jl}{2}\right) \right] ,  \label{i8}
\end{equation}
which, by its turn, can be seen as the inner product of two vectors $\left\{
\psi _{k}\exp \left[ -\frac{2\pi i}{N}jk\right] \right\} $ and $\left\{ \psi
_{k-l}\exp \left[ -\frac{\pi i}{N}jl\right] \right\} $ of unity length. By
the Schwarz inequality it is clear than that $\left| \rho _{s}(j,l)\right|
^{2}\leq 1,$ and from properties of the discrete Fourier transform, one can
also conclude that 
\begin{equation}
\left( \rho _{w}(m,n)\right) ^{2}\leq 1.  \label{p3}
\end{equation}
Now, using the property \cite{ruga} 
\begin{equation}
Tr[\hat{O}_{1}\hat{O}_{2}]=\frac{1}{N}\sum_{m,n}O_{1}(m,n)O_{2}(m,n),
\label{p3.5}
\end{equation}
then 
\begin{equation}
Tr[\left( |\psi \rangle \langle \psi |\right) ^{2}]=\frac{1}{N}%
\sum_{m,n}\left( \rho _{w}(m,n)\right) ^{2},  \label{p4}
\end{equation}
which leads to 
\begin{equation}
1=\frac{1}{N}\sum_{m,n}\left( \rho _{w}(m,n)\right) ^{2},  \label{p5}
\end{equation}
and considering inequality (\ref{p3}) we conclude that the discrete Wigner
function must be different from zero in at least on $N$ sites in the
discrete phase space.
\end{itemize}

\section{The continuum limit in phase space}

The continuum limit of an operator representative in phase space is to be
seen as its behaviour in the infinite dimensional/continuum limit. We now
follow a procedure similar to that of \cite{eu}.

\subsection{Cartesian coordinates}

We start from the discrete space operator basis elements, 
\begin{equation}
G(j,l)=\frac 1N\sum_{m,n=-h}^hU^mV^n\exp \left( \frac{i\pi mn}N\right) \exp %
\left[ -\frac{2\pi i}N\left( mj+nl\right) \right] ,  \label{36}
\end{equation}
were we omit the modular phase since we will restrict ourselves to sums in
the interval $[-h,h].$ Then we introduce the scaling parameter

\begin{equation}
\epsilon =\sqrt{\frac{2\pi }{N}},  \label{7}
\end{equation}
which will become infinitesimal as $N\rightarrow \infty $. We also introduce
two Hermitian operators $\{P,Q\},$ 
\begin{equation}
P=\sum_{j=-\frac{N-1}{2}}^{\frac{N-1}{2}}j\epsilon ^{\delta
}p_{0}|v_{j}\rangle \langle v_{j}|\qquad Q=\sum_{j^{\prime }=-\frac{N-1}{2}%
}^{\frac{N-1}{2}}j^{\prime }\epsilon ^{2-\delta }q_{0}|u_{j^{\prime
}}\rangle \langle u_{j^{\prime }}|,  \label{29}
\end{equation}
constructed out of the projectors of the eigenstates of $U$ and $V.$ Again, $%
\delta $ is a free parameter which might assume any value in the open
interval $(0,2)$. $\{p_{0},q_{0}\}$ are real parameters that might carry
units of momentum and position, respectively, and $\epsilon ^{\delta }p_{0}$
and $\epsilon ^{2-\delta }q_{0}$ are the distance between successive
eigenvalues of the $P$ and $Q$ operators. With the help of these, we can
rewrite the Schwinger operators as 
\begin{equation}
V=\exp \left[ \frac{i\epsilon ^{2-\delta }P}{p_{0}}\right] \qquad U=\exp %
\left[ \frac{i\epsilon ^{\delta }Q}{q_{0}}\right] .  \label{28}
\end{equation}
and perform the change of variables 
\begin{eqnarray}
q &=&q_{0}\epsilon ^{2-\delta }j\text{ \qquad }p=p_{0}\epsilon ^{\delta }l 
\nonumber  \label{30} \\
u &=&p_{0}\epsilon ^{\delta }m\text{ \qquad }v=-q_{0}\epsilon ^{2-\delta }n.
\label{30}
\end{eqnarray}
With that, we arrive at a new operator basis elements that do not
explicitely depend on $\delta $, but, at the same time, the operators $U$
and $V$ carry a particular $\epsilon $ dependence, defined by the particular
choice of $\delta $, namely 
\begin{equation}
G(p,q)=\frac{1}{q_{0}p_{0}\epsilon ^{2}N}\sum_{u,v=-h}^{h}\Delta u\Delta
v\exp \left[ \frac{iuQ}{p_{0}q_{0}}\right] \exp \left[ -\frac{ivP}{p_{0}q_{0}%
}\right] \exp \left( -\frac{i}{2p_{0}q_{0}}uv\right) \exp \left[ -\frac{i}{%
p_{0}q_{0}}\left( qu-pv\right) \right] .  \label{30.5}
\end{equation}
If we take the limit $N\rightarrow \infty ,$ it is clear that we can
consider $\Delta u\rightarrow du$ and $\Delta v\rightarrow dv,$ yielding 
\begin{equation}
G(p,q)=\frac{1}{2\pi q_{0}p_{0}}\int_{-\infty }^{\infty }\int_{-\infty
}^{\infty }dudv\exp \left[ \frac{iu(Q-q-v/2)}{p_{0}q_{0}}\right] \exp \left[
-\frac{iv(P-p)}{p_{0}q_{0}}\right] .  \label{30.6}
\end{equation}
As we know from \cite{eu} that in this limit we recover the usual results
for position and momentum once $p_{0}q_{0}=\hbar $, we use the identity 
\begin{equation}
|q\rangle \langle q|=\frac{1}{2\pi \hbar }\int_{-\infty }^{\infty }dx\exp %
\left[ \frac{ix(Q-q)}{\hbar }\right] ,  \label{30.7}
\end{equation}
and obtain 
\begin{equation}
G(p,q)=\frac{1}{2\pi \hbar }\int_{-\infty }^{\infty }dv|q+v/2\rangle \langle
q+v/2|\exp \left[ -\frac{iv(P-p)}{\hbar }\right]   \label{30.8}
\end{equation}
\begin{equation}
G(p,q)=\frac{1}{2\pi \hbar }\int_{-\infty }^{\infty }dv|q+v/2\rangle \langle
q-v/2|\exp \left[ \frac{ivp}{\hbar }\right] ,  \label{30.9}
\end{equation}
which is exactly the form of the Weyl-Wigner basis elements $\Delta (p,q).$
It is interesting to see that, as in the state space description, the
parameter $\delta $ doesn't affect the final result since, in this case (any 
$\delta \in (0,2)$), the basis elements do not depend on it at all, but $U$
and $V$ depend on $\epsilon $. It is now a trivial matter to prove that the
decomposition coefficients are well behaved in the limit and also go to the
Weyl-Wigner coefficients. From this we see that the whole mapping scheme is
recovered. This result was already achieved for the particular case $\delta
=1$ in \cite{ruga}, where the limiting process which leads to the Moyal
bracket was also discussed. Moreover, it has to be stressed that, starting
from the continuous family of unitary operators, Eq. (\ref{28}), and
realizing \ the independence of the basis elements on $\delta $, the
Weyl-Wigner basis elements are overdetermined in the limiting process,
since, for any $\delta \in (0,2)$ pair of operators, we always get the same
final expression. This means that for the continuous family of unitary
operators (except for $\delta =0$ or $\delta =2$), as proposed, the
continuum limit is the Weyl-Wigner operator basis.

From these results one immediately concludes that the discrete Wigner
function has the ordinary Wigner function as its continuum limit, in the
sense discussed above. As we already stated, most properties of the usual
Wigner function are originally present in the discrete one, and come out as
the continuum limit of the latter.

In the discrete case we have seen that the Wigner function must be different
from zero in at least $N$ sites in phase space. It is obvious that the same
procedure which led to this result would lead to the well known property of
the usual Wigner function that it must be different from zero in a region of
the phase space of area at least $\hbar $. This discussion illustrates
somewhat quantitatively how the quantum effects become more and more drastic
as the dimensionality $N$ gets smaller.

\subsection{Angular coordinates}

Following on our analogy with what was done in \cite{eu}, we choose now the
parameter $\delta $ in the extreme situation $\delta =0$. We expect now to
obtain a phase space formalism which is consistent with angular coordinates.
We start again from our discrete operator space basis elements, Eq.(\ref{36}%
), 
\[
G(j^{\prime },l^{\prime })=\frac{1}{N}\sum_{m^{\prime },n^{\prime
}=-h}^{h}U^{m^{\prime }}V^{n^{\prime }}\exp \left( \frac{i\pi m^{\prime
}n^{\prime }}{N}\right) \exp \left[ -\frac{2\pi i}{N}\left( m^{\prime
}j^{\prime }+n^{\prime }l^{\prime }\right) \right] ,
\]
Rewriting the Schwinger operators as above, but with $\delta =0$, we now
would have 
\begin{equation}
M=\sum_{j=-\frac{N-1}{2}}^{\frac{N-1}{2}}jm_{0}|v_{j}\rangle \langle
v_{j}|\qquad \text{and \ \ \ }\Theta =\sum_{j^{\prime }=-\frac{N-1}{2}}^{%
\frac{N-1}{2}}j^{\prime }\epsilon ^{2}\theta _{0}|u_{j^{\prime }}\rangle
\langle u_{j^{\prime }}|,  \label{32}
\end{equation}
leading to 
\begin{equation}
V=\exp \left[ \frac{i\epsilon ^{2}M}{m_{0}}\right] \qquad \text{and \ \ \ \ }%
U=\exp \left[ \frac{i\Theta }{\theta _{0}}\right] ,  \label{33}
\end{equation}
so that only $V$ depends now on $\epsilon $, and changing the variables as 
\begin{eqnarray}
\theta  &=&\theta _{0}\epsilon ^{2}j^{\prime }\text{ \qquad }%
l=l_{0}l^{\prime }  \nonumber  \label{30} \\
m &=&m_{0}m^{\prime }\text{ \qquad }\alpha =-\theta _{0}\epsilon
^{2}n^{\prime },  \label{34}
\end{eqnarray}
we have for the basis elements 
\begin{eqnarray}
G(\alpha ,l) &=&-\frac{1}{2\pi \theta _{0}}\sum_{m=-m_{0}h}^{m_{0}h}\sum_{%
\alpha =(\pi -\frac{\pi }{N})\theta _{0}}^{(-\pi +\frac{\pi }{N})\theta
_{0}}\Delta \theta \exp \left[ \frac{im\Theta }{m_{0}\theta _{0}}\right]
\exp \left[ -\frac{i\alpha M}{m_{0}\theta _{0}}\right]   \nonumber \\
&&\exp \left( -\frac{im\alpha }{2m_{0}\theta _{0}}\right) \exp \left[ -\frac{%
i}{m_{0}\theta _{0}}\left( m\theta -l\alpha \right) \right] .  \label{36b}
\end{eqnarray}
Performing again the limit $N\rightarrow \infty $, the angle variables
become continuous and we have 
\begin{equation}
G(\alpha ,l)=\frac{1}{2\pi \theta _{0}}\sum_{m=-\infty }^{\infty }\int_{-\pi
\theta _{0}}^{\pi \theta _{0}}d\alpha \exp \left[ im(\Theta -\theta -\frac{%
\alpha }{2})\right] \exp \left[ -\frac{i\alpha (M-l)}{m_{0}\theta _{0}}%
\right] .  \label{35}
\end{equation}
The sum over $m$ is the projector in angle space ($\theta _{0}$ is set to 1,
so the angle units are radians, and $m_{0}\theta _{0}$ is set to $\hbar $),
and 
\begin{equation}
G(\alpha ,l)=\frac{1}{2\pi }\int_{-\pi }^{\pi }d\alpha ||\theta +\frac{%
\alpha }{2}\rangle \langle \theta +\frac{\alpha }{2}|\exp \left[ -\frac{%
i\alpha (M-l)}{\hbar }\right] .  \label{36c}
\end{equation}
so that, with the use Eq.(\ref{32}), we achieve the result 
\begin{equation}
G(\alpha ,l)=\frac{1}{2\pi }\int_{-\pi }^{\pi }d\alpha |\theta +\frac{\alpha 
}{2}\rangle \langle \theta -\frac{\alpha }{2}|\exp \left[ \frac{il\alpha }{%
\hbar }\right] ,  \label{37}
\end{equation}
that is precisely the result of references \cite{muk,biza}. We remark that
we have no need to worry about the periodicities in the angle variable as
our angular states are bounded to the $[-\pi ,\pi )$ interval by {\it %
definition,} and our notation has $%
\mathop{\rm mod}%
N$ periodicity ($%
\mathop{\rm mod}%
2\pi $ in the continuum limit) by construction\cite{eu}. It would seem at
first glance that the continuum interval is $[-\pi ,\pi ],$ but that is not
the case as it can be seem from the original discrete results that the
states in the extremes of the interval are {\em not} the same. We understand
that, once the basis elements are recovered, the whole mapping procedure is
recovered.

Again, all properties of the angular Wigner function can be obtained from
its discrete counterpart by the limiting process above. It must be stated
however that in a lot of cases it turns out to be easier to work with the
discrete rather than in the angular case. That is particularly true in the
obtention of the angular counterpart of Eq.(\ref{p3}), which in the angular
case doesn't lead to a condition involving a minimal area unit in phase
space due to the very nature of the angular phase space.

It is interesting to note that what was considered to be {\em conditions}
for the existence of the Wigner function in \cite{muk,biza} are derived as
properties of it in the present scheme.

\subsection{Mapping of the Pegg and Barnett operators}

The number and phase operators of Pegg and Barnett can be immediately mapped
on the discrete phase space. In fact, we exactly reproduce the PB scheme if
we rename the \ $M$ operator of Eq.(\ref{33}) by $N$ and include a reference
angle in the definition of \ $\Theta $ (which must be an integer multiple of 
$\frac{2\pi }{N}$). The phase space representatives of these operators,
through direct use of Eq.(\ref{13}), are seen to be 
\begin{equation}
N(m,n)=n,\qquad \Theta (m,n)=\theta _{ref}+\frac{2\pi }{N}m,  \label{38}
\end{equation}
with obvious continuum limits.

\section{Conclusions}

Motivated by the results of part I, we looked for a phase space discussion
of the limits which connect discrete, angular and Cartesian coordinates. It
became then clear that the Weyl-Wigner formalism, in both position-momentum 
{\em and} angle-angular momentum cases, can be seen as limiting elements of
a discrete phase space formalism. The angle-angular momentum case is seen to
be in deep connection with the Pegg and Barnett approach to the phase
problem, while the Weyl-Wigner operator basis is reobtained for all the
cases for which the parameter governing the unitary operators is different
from zero; in this sense the Weyl-Wigner basis is overdetermined in the
limiting process. An interesting by-product of this discussion is the
analysis of the Wigner function, which reproduced the conditions imposed on
the angular Wigner function in \cite{muk,biza}.

With all that in mind, one is compelled to regard this as a kind of
standard, or rather `natural' approach to phase space in quantum mechanics.
The basic feature that pertains to all three versions of the formalism is
that one constructs a basis in operator space out of the Fourier transform
of the shifting operators. A one-to-one correspondence then ensures the
existence of a mapping between abstract operators and functions in phase
space.

\end{document}